\documentclass[9pt,twocolumn,twoside]{osajnl}

\journal{ao} 

\setboolean{shortarticle}{true}

\usepackage[varg]{txfonts}   
\usepackage{booktabs}
\usepackage{array} 
\usepackage{graphicx}
\usepackage[caption=false]{subfig}
\usepackage{lipsum} 
\usepackage{cuted}
\usepackage{flushend}
\usepackage{hyperref}
\newcolumntype{L}[1]{>{\raggedright\let\newline\\\arraybackslash\hspace{0pt}}m{#1}}
\newcolumntype{C}[1]{>{\centering\let\newline\\\arraybackslash\hspace{0pt}}m{#1}}
\newcolumntype{R}[1]{>{\raggedleft\let\newline\\\arraybackslash\hspace{0pt}}m{#1}}
\graphicspath{{graphics/}{graphics/arch/}{Graphics/}{./}} 
\usepackage{xcolor}

\newcommand{\un}[1]{\ensuremath{\,{\rm #1}}}

\newcommand {\beq} {\begin{equation}}
\newcommand {\eeq} {\end{equation}}

\title{ A Compact Millimeter-Wavelength Fourier-Transform Spectrometer}

\author[1,2,*]{Zhaodi Pan}
\author[1,2]{Mira Liu}
\author[3] {Ritoban Basu Thakur}
\author[1,4,5] {Bradford A. Benson}
\author[6,7] {Dale J. Fixsen}
\author[8]{Hazal Goksu}
\author[2]{Eleanor Rath}
\author[1,2,4,9,*] {Stephan S. Meyer}

\affil[1]{Kavli Institute for Cosmological Physics, University of Chicago, Chicago, 5640 S. Ellis Ave., IL 60637, USA}
\affil[2]{Department of Physics, University of Chicago, 5640 S. Ellis Ave., Chicago, IL 60637, USA}

\affil[3]{California Institute of Technology, MC 249-17, Pasadena, CA 91125, USA}
\affil[4]{Department of Astronomy and Astrophysics, University of Chicago, 5640 S. Ellis Ave., Chicago, IL 60637, USA}
\affil[5]{Fermi National Accelerator Laboratory, MS209, P.O. Box 500, Batavia, IL 60510}
\affil[6]{Code 665, Goddard Space Flight Center, Greenbelt, MD 20771, USA}
\affil[7]{University of Maryland, College Park, MD 20742, USA}
\affil[8]{Department of Physics, ETH Zürich, CH-8093 Zürich, Switzerland}
\affil[9]{Enrico Fermi Institute, University of Chicago, 5640 S. Ellis Ave., Chicago, IL 60637, USA}

\affil[*]{Corresponding authors: meyer@uchicago.edu (Stephan Meyer), panz@uchicago.edu (Zhaodi Pan).}

\ociscodes{To be added.}

\begin{abstract}
We have constructed a Fourier-transform spectrometer (FTS) operating between 50 and 330~GHz with minimum volume ($\mathbf{355\times 260\times 64}$~mm) and weight (13~lbs) while maximizing optical throughput ($\mathbf{100}~\mathrm{\mathbf{mm}^2}$sr) and optimizing the spectral resolution (4~GHz). This FTS is designed as a polarizing Martin-Puplett interferometer with unobstructed input and output in which both input polarizations undergo interference.  The instrument construction is simple with mirrors milled on the box walls and one motorized stage as the single moving element. We characterize the performance of the FTS, compare the measurements to an optical simulation, and discuss features that relate to details of the FTS design. The simulation is also used to determine the tolerance of optical alignments for the required specifications. We detail the FTS mechanical design and provide the control software as well as the analysis code online.

\end{abstract}

\begin{document}

\maketitle

\section{Introduction}
\label{s1}

We have constructed a Fourier Transform Spectrometer (FTS) operating at mm and sub-mm wavelengths as a prototype similar to the instrument designed for a NASA MIDEX mission (PIXIE) \cite{kogut11}. This instrument was used to characterize the end-to-end spectral response of a kilo-pixel radiometer constructed for use on the South Pole Telescope (SPT)\cite{carlstrom201110}. It was also used to cross-check detector spectral response measurements for Keck Array \cite{staniszewski2012keck}. 

We modified the polarizing Martin-Puplett FTS design of the FIRAS instrument \cite{martin70, firas} to optimize for a smaller and lighter instrument with maximum optical throughput, while preserving the high efficiency, dual polarization, and polarization preserving properties. The design specifications are the total optical throughput, the maximum and minimum frequencies of operation, and the spectral resolution at the maximum frequency. The specifications constrain the size and roughness of the optics, the maximum mirror motion, and the minimum size of optical elements. These considerations are combined with the desire for a light, compact and simple system with no optical adjustments.

We characterize the FTS with a bolometric detector and couple it to both incoherent blackbody sources and single-mode spectrally unresolved sources at three frequencies. The single-mode source is mounted on an X-Y stage in the input focal plane so that its position can be modified. Both sources have modulators to provide a chopped signal.  The measurements are compared to a ray-trace simulation. 

In this paper, we describe the FTS design in Section \ref{section:design}; the experimental procedures and software in Section \ref{section:setup}; the measurements and characterization in Section \ref{section:measurements}; and compare the measurements to the simulation in Section \ref{section:simulation}. In Section \ref{section:applications}, we describe the applications of the FTS, including characterizing the spectral response of the SPT-3G camera.

\section{FTS Design}
\label{section:design}
\subsection{Design overview}
\label{section:overview}
\begin{figure}[!ht]
  \centering
  \includegraphics[width=1\linewidth]{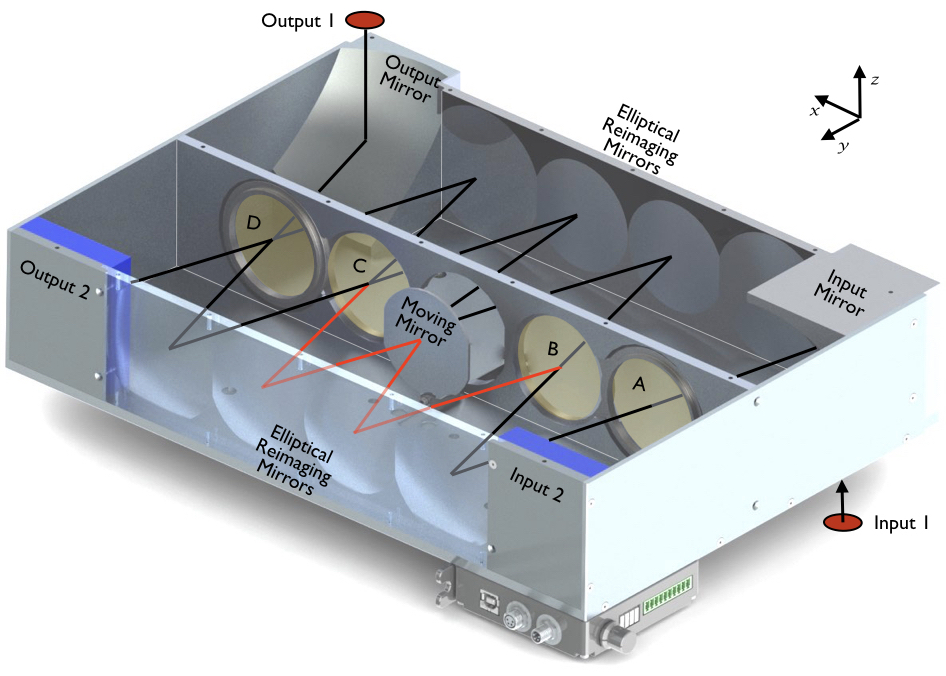}
  \caption{A rendering of a CAD model of the FTS with the near side plate shown transparent.  The polarizers are labeled A through D. On side 1, ellipsoidal input and output mirrors focus the beam 90~deg out of the plane of the optics onto 17~mm diameter focal regions, shown as red disks and located 38~mm below and above the outer surface of the box. On side 2, in this configuration, the input and output ports are ambient temperature HR25 absorbers \cite{eccosorb_hr} shown in blue. On each side of the box, four ellipsoidal reimaging mirrors are machined into the box sidewall. The central moving mirror is shown at its maximum positive delay position. Lines trace the path of the central ray through the instrument for one polarization (the red one in Fig~\ref{fts_schematics}) entering at Input 1 on the right. For one of the two paths between the two beam splitters the ray is shown in red.} 
  \label{fts_picture}
\end{figure}

The FTS described in this paper is a polarizing Martin-Puplett \cite{martin70} interferometer that has been modified from the typical Michelson configuration to a Mach-Zehnder \cite{zehender, mach} arrangement.  The design has two input ports, which are well separated from the two output ports, and two beam splitters. The typical rooftop mirrors in the Martin-Puplett design, used to rotate the polarization, are unnecessary since the beams splitters can be oriented at 90$^\circ$ relative to each other.  This design requires four identical polarizers rather than the usual two for the classic Martin-Puplett design, however the size of the polarizers is smaller, maintaining a compact design. A CAD model of the FTS is shown in Fig.\ref{fts_picture}. The design parameters for this FTS is a frequency range of 50 to 330~GHz, a resolution of 1~GHz, and throughput of 100~\un{mm^2\,sr}. The resulting instrument is $355\times 260\times 64$~mm in size.

To minimize the size of the instrument and to permit all the internal optics to be machined into the instrument sidewalls the beam is directed at angle relative to the motion of the moving mirror. This angle introduces two non-idealities: 1) the loss of beam as the mirror moves slightly sideways relative to the beam with increasing delay, and 2) the relative displacement of the beams passing on opposite sides of the moving mirror as they reach the detector plane. If displaced far enough, their power adds at the detector rather than interfering.

The ellipsoidal mirrors are configured so that each re-images the previous mirror onto the next. Each mirror is a part of the ellipsoid that has the last and the next mirror centers at the foci. The remaining parameter of the ellipsoids is chosen so that the mirrors just intersect at the inner surface of the box. The first and last mirrors are also ellipsoidal but have the input or output ports at one focus and the next mirror at the other. The input and output ports are placed 38~mm outside the surface of the box. The angle of the beams in the box is $\phi=\cos^{-1}2r/y$ where $r$ is the radius of the mirrors and $y$ is the width of the box (see Fig. \ref{fts_schematics}).

The moving part of the FTS that generates the optical delay is a thin flat mirror on a one-axis stage near the center of the box (referred to as center mirror afterward). For the chief ray in the interferometer, the optical delay is $d =4 y\cos\phi$ where $ y$ is the mechanical displacement of the mirror from the center of the box. Therefore, a mirror displacement is amplified to an optical delay by nearly a factor of 4. 

A center plate spanning the length of the box supports the four polarizers (labeled A through D) and has a precise machined cut-out for the mirror translation. The polarizer wires are gold-coated tungsten wires wound on stainless steel frames which are mounted to the center plate. The polarizing wires of grids B and C are oriented 90$^\circ$ to each other, and A and D are oriented 45$^\circ$ to B and C from the point-of-view of the beam. Note the beam is at angle $\phi$ relative to the $y$ axis (Fig. \ref{fts_schematics}). Switching the relative orientation of A and D (or B and C) from being parallel to being orthogonal (or the other way) switches the symmetric output and antisymmetric output. In our implementation, B has its wires vertical along the $z$ axis and C has its wires along the $x$ axis. 

\subsection{Optics}
\label{section:optics}
\subsubsection{Geometrical Layout}
\label{section:geometry}

\begin{figure}[!ht]
  \centering
  \includegraphics[width=.7\linewidth]{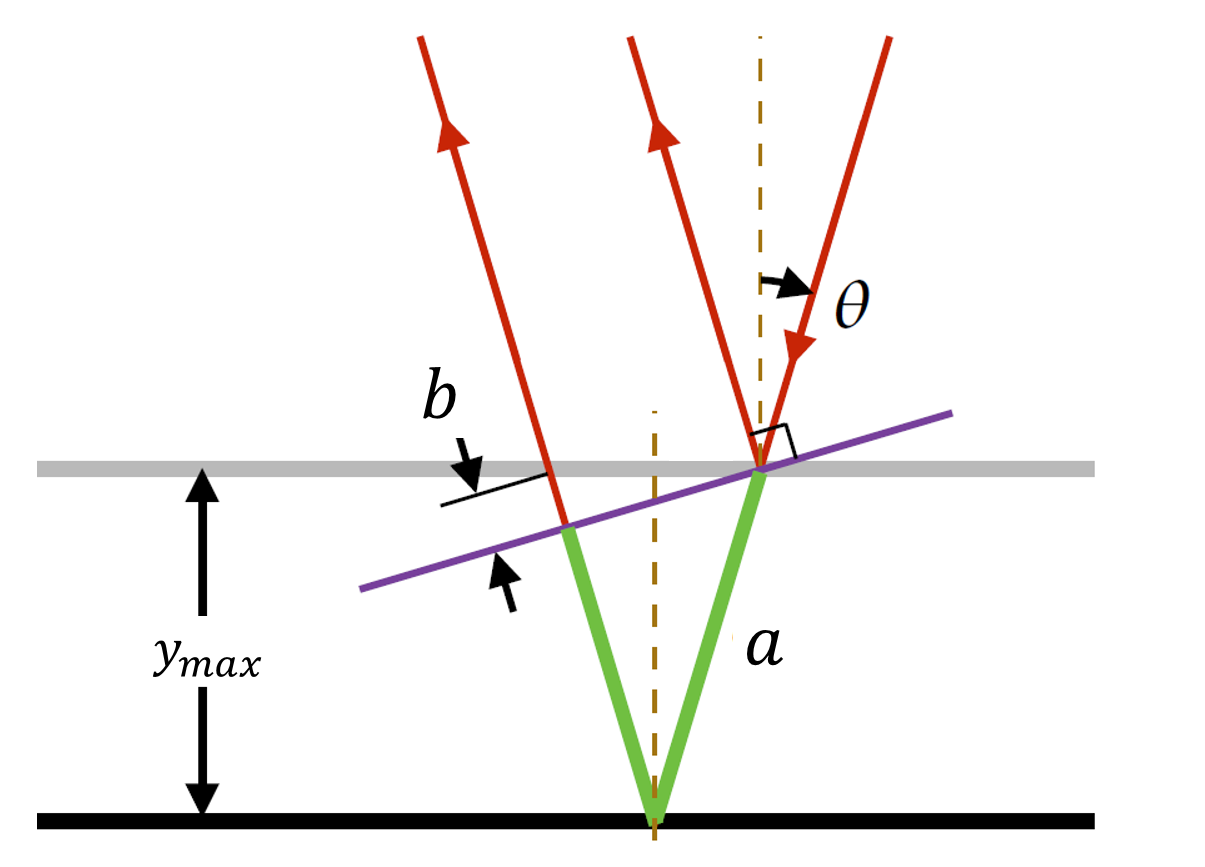}
  \caption{The geometry of the incident light rays when the center mirror is at its maximum displacement $y_{max}$. The solid black and gray lines are the mirror planes at zero and maximum displacements. The red and green lines are the light paths of a light ray incident on the mirror at $\theta$ when the mirror is at zero and maximum optical delays. The purple line indicates the plane of the wavefront.  The finite solid angle of the beam results in a path length difference between rays at the center of the beam where $\theta=0$ and  at the edge of the beam where $\theta_{\rm max}\sim\sqrt{\Omega/\pi}$.  }
  \label{geometry}
\end{figure}

The essential advantages of our FTS design shown in Fig~\ref{fts_picture} are the high density of beams, the separation of input and output optics, and the simplicity of optics in the sidewalls. This layout also has the advantage of a single small moving mirror with beams impinging on both sides. Once this geometry is selected, the desired spectral resolution and frequency range define the geometry of the system. The resolution is limited by the solid angle of the beam at the beam splitter, $\Omega$, because high angle rays have a different optical delay than the central ray \cite{chamberlain} and thus decohere with higher delay. For a given resolution, the limited solid angle together with instrument throughput requirement defines the size of the beam splitters and the mirrors. Chamberlain \cite{chamberlain} calculates an analytic approximation of the decoherence for a tophat distribution of angles and integrating over all beams. For this design, we set the added delay ($b$ in Fig~\ref{geometry}) for the largest angle beam to be one wavelength at the maximum delay for the highest design frequency.  This constrains $b=\lambda_{\rm min}=c/f_{\rm max}$, where $f_{\rm max}$ is the highest design frequency. The decoherence constraint implies that the beam center and the beam edge have a phase difference of $2\pi$ at the minimum wavelength $\lambda_{min}$ when the center mirror is at its maximum displacement reducing the contrast to near zero.

\begin{figure}[!ht]
  \centering
  \includegraphics[width=1\linewidth]{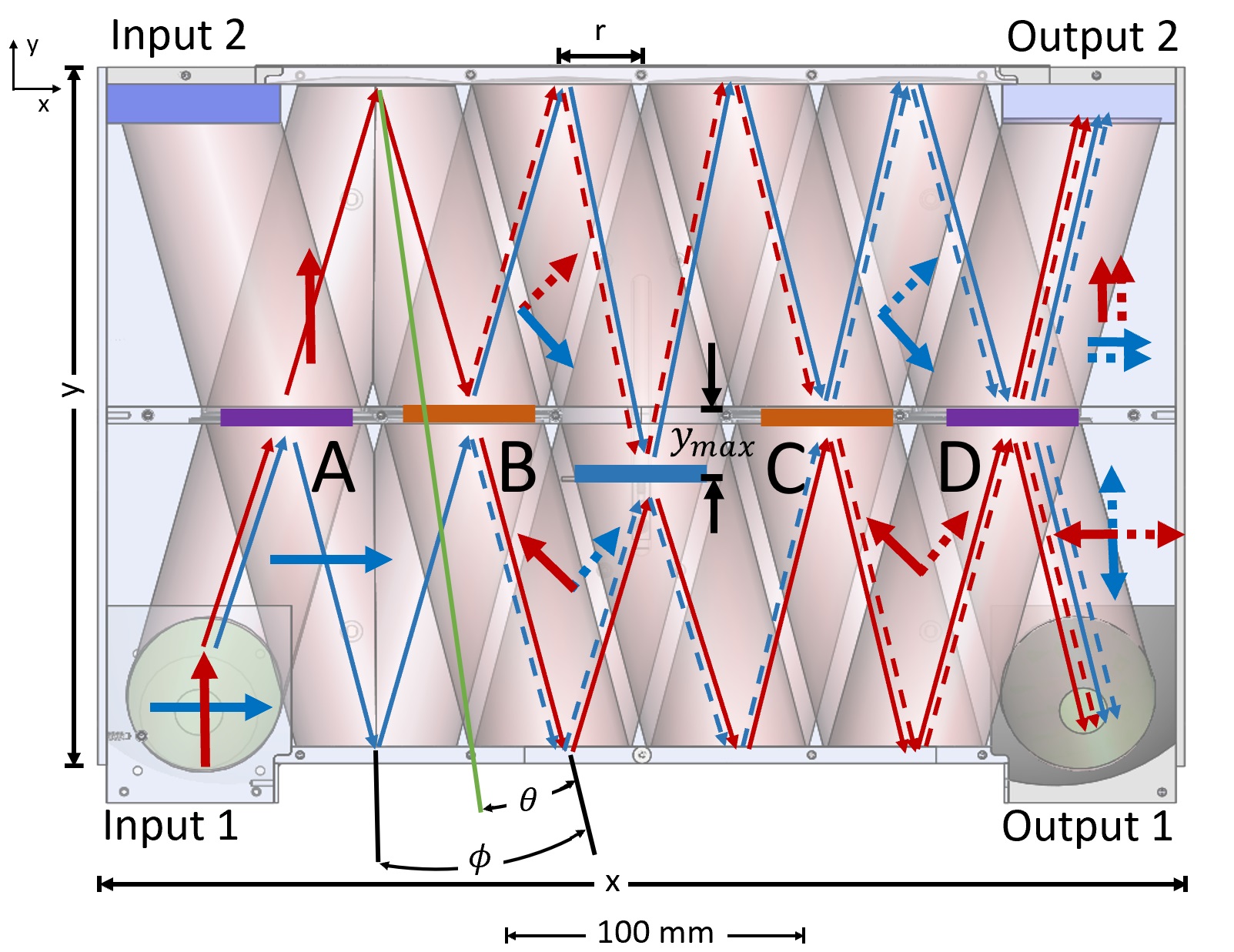}
  \caption{Schematic of the FTS optical geometry and polarization. The input and output ports are labeled. The polarizers are located on the center septum, the reflection mirrors on the top and bottom of the figure. The center mirror is in blue. In this figure, we also track light paths and polarization directions from Input 1. Thin arrows represent optical paths, and thick arrows the polarization directions. Solid and dashed lines are different polarization components. $y_{\rm max}$ is the maximum displacement of the center mirror, $r$ is the radius of the beam at the mirrors, $\phi$ is the angle the beams make with the $y$ axis, and $\theta$ is the opening angle used to calculate the solid angle at the beam splitter.}
  \label{fts_schematics}
\end{figure}

 The size of the FTS is calculated from the design parameters using the constraint from Fig. \ref{geometry} and assuming the effect of the incident angle of the beam ($\phi$ in Fig. \ref{fts_schematics}) is negligible. The maximum optical delay $d_{max}$ then is 4 times the center mirror's maximum displacement $y_{max}$, and is set by the desired resolution $\Delta f$ \cite{Persky95}:

\begin{equation}
d_{max}= 4y_{max} = c/\Delta f .
\label{resolution}
\end{equation}

From Fig. \ref{geometry}, we can calculate the path difference between the beam center (the vertically incident rays) and the beam edge (the rays incident at angle $\theta$ ), which is $\delta = 2y_{max}-2a+b$, where $a$ and $b$ are defined in the figure. $a$ and $b$ can be related to $y_{max}$ and $\theta$ by $a=y_{max}/\cos\theta$ and $b= 2y_{max}\tan\theta \sin\theta$. With these geometric relations, $\delta$ can now be written as a function of $y_{max}$ and $\theta$ as follows:

\begin{equation}
\delta=2y_{max}\left(1-1/\cos\theta+\tan\theta\sin\theta \right) = y_{max}(2\sin(\theta/2))^2
\label{eqdelta}
\end{equation}

Eq. (\ref{resolution}), Eq. (\ref{eqdelta}), and the decoherence constraint $\delta=c/f_{max}$ can be combined to give the dependence of $\theta$ on  $f_{max}$ and  $\Delta f$:

\begin{equation}
\theta=2\arcsin(\sqrt{\Delta f/f_{max}}).
\label{theta}
\end{equation}

Now we calculate the beam splitter diameter $r$. The required throughput of the system $A\Omega$ is the product of the area $A=\pi r^2$ and the solid angle $\Omega= 4\pi(\sin(\theta/2))^2$, and can be written as $A\Omega=(2\pi r \sin(\theta/2))^2$. This combined with Eq.(\ref{theta}) gives:

\begin{equation}
r= \frac {1}{2\pi} \sqrt{\frac{A\Omega f_{max}}{\Delta f}}
\end{equation}

The length, width, and height of the FTS box can be related to $r$ by $x=12r$, $y=2r/\tan(2\theta)$, and $z=2r$ (Fig. \ref{fts_schematics}). At this point, all size parameters are expressed as functions of the three design parameters $f_{max}$, $\Delta f$, and $A\Omega$. For the system tested here, the design parameters are $f_{max}=330$~GHz, $\Delta f=1$~GHz, and $A\Omega=100~\mathrm{mm}^2$sr. The resulting design has $\theta= 6.3$~deg, $r=29$~mm, $x=347$~mm, $y=260$~mm, and $z=58$~mm. The size of a fabricated FTS is slightly larger ($355\times 260 \times 64$~mm), with finite thicknesses of the metal walls and metal edges.

\subsubsection{Mirror Figures}
\label{section:mirror_figure}
Each mirror in the system is a small section of a prolate ellipsoid of revolution about a line with two foci. One focus is the center of the previous mirror and the second at the subsequent mirror. For the input and output mirrors one focus is at the center of the source plane or the output plane, and the other focus is at the next mirror. With the two foci fixed, only one further parameter is needed. For this optical design, the position of the intersection of the central ray with the mirror surface is convenient. Once the radius of the mirrors and the width of the box is determined, the y position of the center of the four mirrors in the sidewall is chosen so that the intersection of the inner surface of the wall and the ellipsoid has the required diameter. 

The choice of input and output focal position depends on the use for the FTS. In the configuration of the FTS used for measuring the SPT cryostat, the output mirror was rotated and the focal point positioned so that the beam could be injected into the cryostat. For the tests described here, the configuration is as shown in Fig~\ref{fts_picture}.

\subsection{Specifications for hardware}
\label{section:hardware}
 The FTS metal parts are computer numeric control (CNC) machined from Aluminum 6061, with a machined precision of $\pm 0.12$~mm. The surface roughness of the machined mirrors is 3.2~$\mu$m Ra. The polarizers used are wire grids made of 25-$\mu$m-diameter gold-plated tungsten wires spaced at 100-$\mu$m intervals. The center mirror is moved using a linear driver, Zaber LSM050B model \cite{zaber} with a spatial accuracy of 25$~\mu$m and a speed resolution of 0.9$~\mu$m/s. The linear driver's speed can vary between 0.9~$\mu$m/s and 29~mm/s, and its travel range is 50.8~mm.  To reduce stray reflections, the inside of the FTS box is coated with Eccosorb HR10 \cite{eccosorb_hr}.

\section{Setup and operation}
\label{section:setup}
\subsection{Test configurations}
\label{section:operation}

\begin{figure}[!ht]
  \centering
  \includegraphics[width=.6\linewidth]{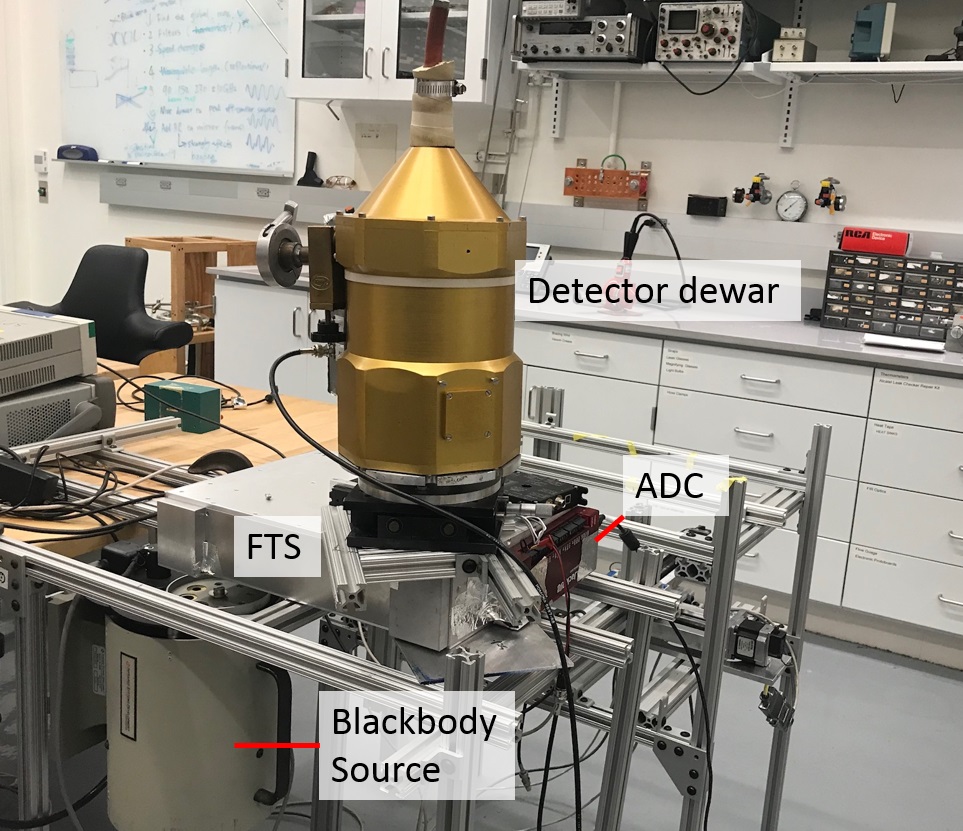}
  \caption{A photo of the FTS test setup. The FTS is the silver box in the middle of the picture. The input port is coupled to a 1300~K blackbody source and the output port is coupled to a monolithic silicon bolometer cooled by the helium dewar in gold. }
  \label{operation}
\end{figure}

To characterize the instrument we have made measurements in a number of configurations. The detector for these tests is a monolithic silicon bolometer \cite{downey84} operated at 4.2~K with throughput of 50~\un{mm^2\,sr} and a sensitive frequency range of 50-600~GHz.  The second output of the FTS is not used and has a room temperature absorber as shown in Fig.~\ref{fts_picture}.  Tests were conducted with two different sources on the antisymmetric input of the interferometer:  1) an IR-563 1300~K blackbody \cite{ir_563} with an adjustable aperture from 2.5~mm to 25.4~mm diameter or 2) a single-mode narrow-band source consisting of a Gunn oscillator \cite{carlstrom85} at 90~GHz, 144~GHz, or 295 GHz (the latter is frequency tripled from the 98~GHz source). The single-mode source was mounted in an X-Y stage and could be moved around the focal plane of the input mirror. Both sources had output beams which fully illuminated the first input optic. The source port on the symmetric side of the interferometer is a room temperature absorber. The placement of the hot source on the antisymmetric side gives a negative interferogram. Both sources were outfitted with a chopper so their output could be 100\% modulated (chopped) over a range of chopper frequencies. Fig. \ref{operation} shows the physical setup during  operation, where the FTS is coupled to a 1300~K blackbody at the input port and a bolometer (within the gold-colored helium detector dewar) located at the output port.

Characterizing the FTS with the blackbody source tests the frequency range and broadband performance, whereas the narrow-band Gunn oscillators measures the frequency accuracy and spectral resolution. For the single-mode source, the X-Y stage is used to move the source over the entire input focal plane. The resolution of the system to an extended source with \'{e}tendu greater then $\lambda^2$ is determined by averaging together interferograms with a grid of single-mode source positions.  In the following sections we will discuss and compare measurements with respect to design targets.

\subsection{Operation}
\label{section:software}

The only moving part of the FTS is a motorized mirror, which generates the optical delay between the two light paths.  The FTS operation first consists of selecting the mirror scan speed and maximum delay.  Digital signals indicate the mirror position, with a constant mirror velocity achieved after a short acceleration phase, and the position of the white light fringe. The control software is publicly available \cite{code}. 

In our FTS configuration, the alignment of the mirrors and polarizers are fixed except for the central moving mirror, whose position needs to be synchronized with the bolometer data. During the FTS characterization, the bolometer output voltage was sampled asynchronously with the mirror motion. To correct for this, the data was oversampled by more than a factor of 5 above the Nyquist frequency (and above the post-detection bandwidth of the detector), and the position was determined by a hardware-derived white light fringe indicator, which was also sampled at the same rate.

\section{Measurements and characterizations}
\label{section:measurements}
\subsection{Interferograms and frequency bands}
\label{section:interferograms}
\begin{figure}[!ht]
  \centering
  \includegraphics[width=1\linewidth]{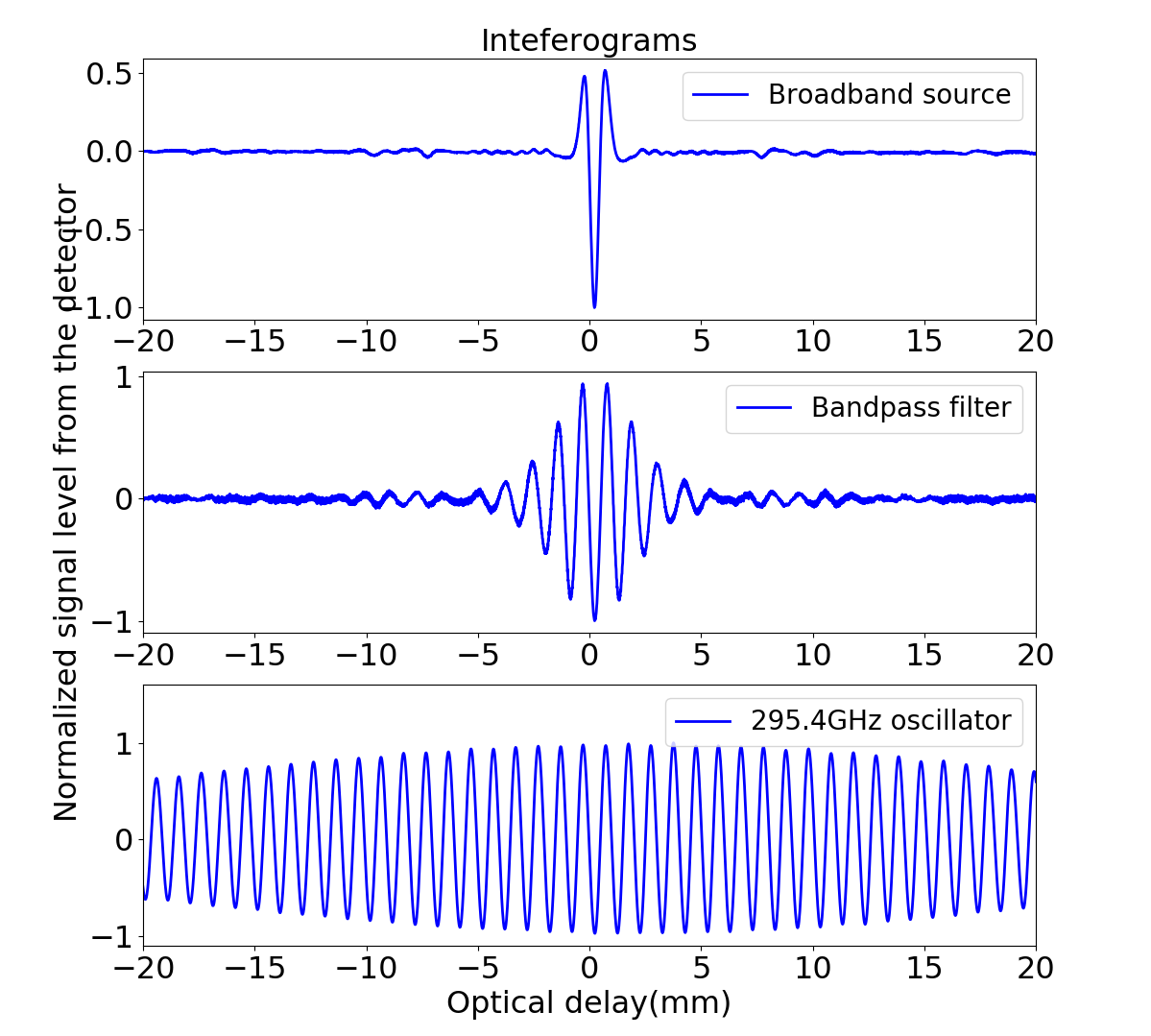}
  \caption{Three example normalized interferograms measured with the FTS central mirror moving at an optical speed of 10~mm/s. The top figure has the FTS configured with a 1300~K source at one input, and an ambient temperature at the other. The only filters used are the absorptive low-pass filters inside the detector dewar, which eliminate high frequency thermal infrared emission.  The second interferogram has a 280~GHz band-pass filter behind the hot thermal source, and the third has a 295~GHz Gunn oscillator at the source input. }
  \label{interferograms}
\end{figure}

\begin{figure}[!ht]
  \centering
  \includegraphics[width=1\linewidth]{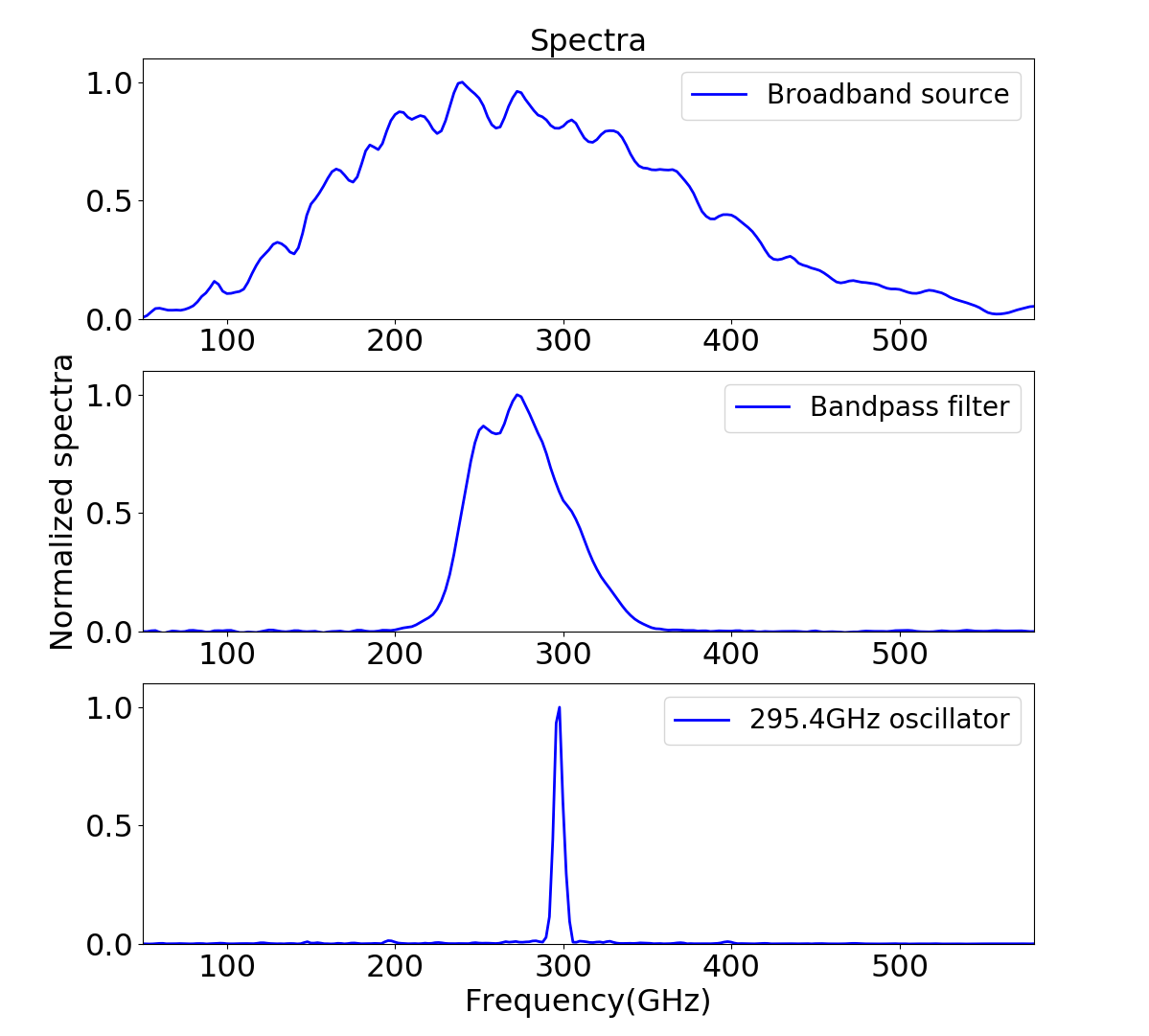}
  \caption{Normalized spectra corresponding to the interferograms of Fig. \ref{interferograms}. }
  \label{spectra}
\end{figure}

Fig~\ref{interferograms} and \ref{spectra} show example interferograms and spectra used to characterize the operating properties, determine the frequency resolution, and measure the spectral accuracy.  Here we tested three sources with very different bandwidths and coherence lengths: a 1300~K broadband blackbody source, the same 1300~K blackbody source paired with a band-defining filter, and a 295~GHz narrow-band tripled Gunn oscillator.  The properties of these sources are known, so we can compare the expected spectra with the measurement. 

The data is a measure of the optical power through the FTS as a function of the optical delay, i.e., the interferogram.  It corresponds to the auto-correlation function of the input radiation, and its Fourier transform is the frequency spectrum (Wiener-Khinchin theorem). The interferogram of the broadband source decays quickly away from zero optical delay. The corresponding spectrum shows the atmospheric absorption spectrum and the detector's response. At low frequencies, the spectrum grows quadratically, following a blackbody spectrum in the Rayleigh-Jeans regime. At high frequencies, the spectrum is shaped by the cryogenic low-pass absorptive filter in receiving detector cryostat. The dip around 557~GHz is water vapor's absorption line. Fig. \ref{interferograms} \textit{middle} is the measured interferogram of the same 1300~K blackbody source with a band-defining filter in the optical path. The input radiation has a narrower band so it decoheres more slowly with an increased optical delay. Fig. \ref{spectra} \textit{middle} is the corresponding frequency band, which is the frequency band in Fig. \ref{interferograms} \textit{top} multiplied by the filter's transmission function. The bottom interferogram with the tripled oscillator shows the response of the instrument to an unresolved source. The tripled Gunn oscillator has a <1~MHz bandwidth centered at 295.4~GHz. The decoherence length is long and is beyond the maximum optical delay. The corresponding spectrum is narrow with a width dominated by the resolution of the FTS and will be discussed in Section \ref{section:measurements}\ref{section:decoherence}. 

The interferograms and the corresponding spectra agree with the source and water absorption line's frequencies. These measurements show that our FTS covers the desired frequency range and is able to measure a wide range of spectra.

\subsection{Instrument characterizations}
\label{section:characterization}

\subsubsection{Characterization overview}
\label{section:characterization_overview}

In Section \ref{section:measurements}\ref{section:interferograms} we discussed test data illustrating the basic function of the FTS. Testing the instrument non-idealities requires analyzing the interferograms quantitatively. A quantitative determination of the FTS properties involves separating the various resolution limiting effects. In addition to the effects of the finite beam solid angle at the beam splitters discussed in Sec.~\ref{section:design}\ref{section:geometry} there are two other effects to be investigated. The first is the simple loss of throughput due to the mirror not moving parallel to the beam by the angle $\phi$. This causes spillover since the beams nearly fill the mirrors at zero delay and increasingly miss the mirror as it moves. This reduces the throughput in a delay-dependent way limiting the instrument resolution. 
The second effect is due to the separation of the recombined beams interfering at the detector as the delay increases. This does not reduce the total throughput but does lower the interference and thus the contrast with delay and the instrument resolution.

The non-idealities above are inevitable because the size limit conflicts other design requirements like the resolution and throughput. To make sure these effects are balanced and within the design limits, we characterized all of them carefully. We present measurements on the transfer efficiency, the modulation contrast, the frequency shift, and the resolution in \ref{section:measurements}\ref{section:efficiency}, \ref{section:measurements}\ref{section:contrast}, \ref{section:measurements}\ref{section:beam_pattern}, and \ref{section:measurements}\ref{section:decoherence}, respectively.

\subsubsection{Transfer efficiency}

\label{section:efficiency}

Not all of the input optical power is able to reach the output ports due to absorption, diffraction, scattering from the surface roughness of the mirrors, and spillover off the edge of the optics. The portion of the optical power that makes its way through the FTS box is defined as the transfer efficiency $\eta$. The transfer efficiency was measured by comparing the response of the detector viewing chopped thermal source through the FTS with a small optical delay with that of the same source illuminating the detector through an reference optical system consisting of only the input and output mirrors. The input mirror or the output mirror has one focus coupled to the detector or the thermal source and the other focus coupled to the other mirror. The two mirrors are spaced by $y/(\cos\phi)$, where $y$ is the width of the box and $\phi$ is the beam's tilt angle, so the two mirrors are at each other's second foci.  The ratio of the two responses is an estimate of the transfer efficiency of the FTS relative to a system with the same input and output geometry but no internal optics. The output power measured at one output of the FTS is multiplied by two to account for the power at the second output port. The measured transfer efficiency is $\eta=92\%\pm 5\%$ and does not change with the input source's frequency spectrum. This meets our requirement because the FTS is able to couple the majority of the input power to its outputs. The transfer efficiency quoted above was measured with the moving mirror close to zero optical delay. The transfer efficiency is lower at higher optical delays because more of the beam is not captured by the moving mirror(Fig. \ref{beamloss} \textit{top}).

 \subsubsection{Modulation contrast}
 \label{section:contrast}
 \begin{figure}[!ht]
  \centering
  \includegraphics[width=1\linewidth]{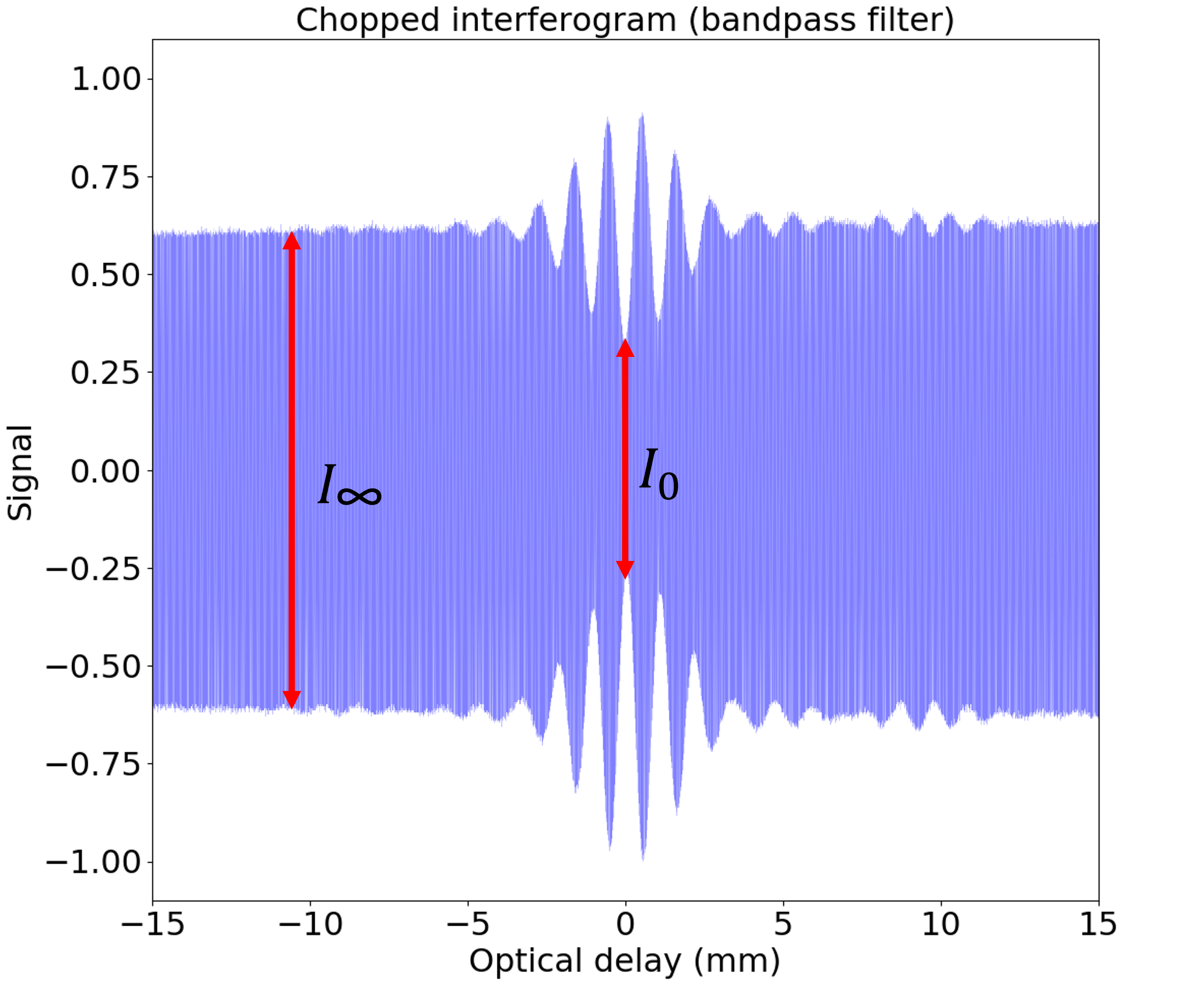}
  \caption{An example of the chopped interferogram.  The blue area is unresolved chopper modulation at a much higher oscillation frequency than the interference pattern, which traces the outer profile of the blue area. The measured chopped depth at the output is proportional to the output intensity. $I_{0}$ and $I_{\infty}$ are intensities at zero and infinite optical delays.  }
  \label{choppedscan}
\end{figure}

The contrast of the FTS is the portion of the transferred radiation that interferes. The contrast can be reduced by several factors, including: non-ideal polarization, non-uniform optical delays for light rays within the same beam, and separation of the recombined interfering beams passing on opposite sides of the mirror at large mirror displacements. The loss of contrast is due to power arriving at the detector, but not interfering completely. The loss is worse at higher source frequencies because the phase non-uniformity of the light rays is larger and the Airy diffraction patterns of the interfering beams are smaller. To quantify these effects, we operated the FTS mirror at very low speed and modulated (chopped) the source. The result is shown in Fig.~\ref{choppedscan}, where we used the band-pass filter. The blue area is the unresolved modulation signal with the chopped signal envelope showing the interferogram. The modulation depth is proportional to the power from the source collected by the detector. An FFT of this chopped signal has the spectrum of the response to the source as AM sidebands of the chopper frequency. The figure shows the modulation depth at large delay, $I_\infty$, and at zero delay, $I_0$ as shown on the figure. Because this is the antisymmetric port, an ideal interferogram would have zero modulation at zero delay. The contrast at small optical delay is defined as $C=I_{0}/I_{\infty}-1$ and depends on frequency. The contrast for the band-defined source in the center plot of Fig.~\ref{interferograms} is $-55\pm 3\%$. When the source has no filter as in the top graph of Fig.~\ref{interferograms}, $C=-33\pm 1\%$, which is expected since the contrast is lower at higher frequencies due to all the effects stated above. The contrast is lower at higher optical delays due to the separation of the recombined beams.

\subsubsection{Frequency shift and input intensity map}
\label{section:beam_pattern}

\begin{figure}[!bht]
  \centering
  \includegraphics[width=\linewidth]{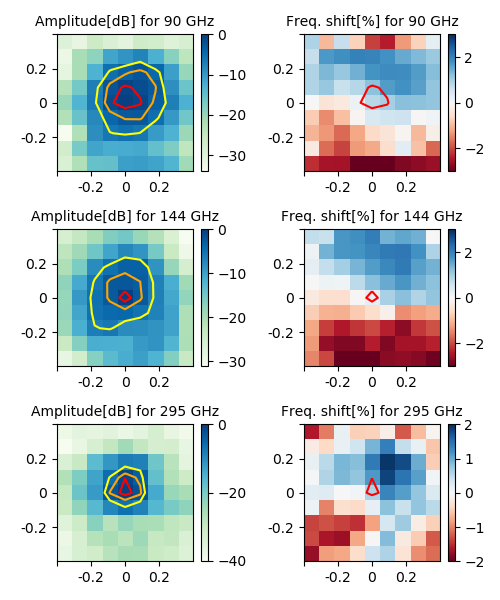}
  \caption{\textit{Top right}: Fractional difference, in percntage, between FTS-determined frequency and the known frequency for a $90.4 \pm 0.1$~GHz Gunn oscillator source as it is moved in a grid across the input focal plane. X and Y axes are coordinates on the input focal plane in inches. Total effective loss of resolution and shift of an extended source is found by weighting these shifts by the intensity map on the top left. \textit{Top left}:Intensity map measured for a $90.4 \pm 0.1$~GHz Gunn oscillator when it is scanned in the same grid as in \textit{top right}. The maximum intensity is normalized to one, and the intensity is plotted in dB with red, orange, and yellow contours indicating -1, -3, and -6~dB (or 79\%, 50\%, and 25\%) levels respectively. \textit{Middle} and \textit{bottom} are similar maps for a $144.3 \pm 0.1$~GHz and a $295.4 \pm 0.1$~GHz Gunn oscillator. The signal decays faster for the 295~GHz source, and the corresponding frequency shift measurement is noisier towards the sides. The -1~dB contours in amplitude are over-plotted in the frequency shift maps. }
  \label{beammap}
\end{figure}

The accurate measurement of the source frequency depends on accurate modeling of the optical delay. An extended source can be thought of as many point sources illuminating the FTS box from different locations on the focal plane. The radiation components from these point sources are transferred at different angles relative to the optical axis and have different optical delays. In our analysis, we use one single equation ($d=4y\cos{\phi}$) to calculate the optical delay based on the motorized actuator's location for all the \'etendu, whereas in reality an integral of the delay over the solid angle would be used. As the beam weighting is not known well,  calibrating the frequency shift and its dependence on the source's position is useful for understanding the frequency accuracy and can be used to estimate and correct for the frequency shift.

The location-dependance of the frequency shift was measured by moving the Gunn oscillator point source across the input focal plane and determining the source's frequency. Fig. \ref{beammap} \textit{middle right} shows an example frequency shift measurement. X and Y are the coordinates of the Gunn oscillator on the input focal plane. The actual frequency of the Gunn oscillator is measured by a spectrum analyzer to be $144.3\pm 0.1$~GHz.  The FTS determined frequency's fractional shift relative to the known frequency varies between $-2.5\%$ and $2.5\%$  across a $\pm0.4\times \pm 0.4$ in area of the input focal plane. The FTS determined frequency is accurate at the (0,0) location and deviates as the source is moved away from the center. The pattern of the frequency shift is sensitive to the alignment of the polarizers and the center mirror (see simulation in Section \ref{section:simulation}).  We performed similar measurements for a 90~GHz and a 295~GHz source and found that the frequency shift patterns are similar. The frequency shift is non-negligible and is caused by assuming a uniform optical delay between the interfering beams. The formula we use for the optical delay ($d=4y\cos\phi$) is accurate for the center position if we ignore the beam divergence. With the beam divergence it is $d=4y\cos{\psi}\cos\phi^{\prime}$ for a single ray, where $\psi$ is the angle relative to the x-y plane, and  $\phi^{\prime}$ is the angle relative to the y-axis in the projected x-y plane. The optical delay for the full beam is the weighted average for all the light rays over the \'etendu.  We explored ways to correct for the inaccurate optical delay in Section \ref{section:simulation}. The frequency shift has two effects, it reduces the resolution of the instrument when using an extended source, and it changes the beam weighted average frequency from what is expected using the simple formula ($d=4y\cos\phi$) for the optical delay. Fig.~\ref{beammap} gives the impression that a measured frequency could be biased by 2.5\%, depending on the exact coupling of the source to the FTS.  However, since the FTS coupling efficiency to a source varies sensitively with position (see Fig.~\ref{beammap} \textit{middle left}), the bias in the measured frequency for a source with an extended area is much less. We calculated the coupling weight map by normalizing the intensity map (Fig. \ref{beammap} \textit{middle left}) so the sum of all pixel weights is one (with pixel values converted to percent from dB). After weighting the frequency shifts (Fig.~\ref{beammap} \textit{middle right}) by the coupling weight map, the measured source frequency by the FTS is 144.4~GHz, which can be compared to the known Gunn oscillator frequency of 144.3$\pm$0.1~GHz.  Therefore our FTS frequency calibration for an extended 144.3~GHz source is accurate at the level of $\sim$0.1 GHz.

The Gunn oscillator amplitude maps on the left of Fig. \ref{beammap} are made in the same way by recording the total modulated power in interferograms taken over the same grid of source position. The measured FWHMs of the intensity maps for the 90, 144, and 295~GHz sources are 0.4~in, 0.3~in, and 0.2~in, respectively. The FWHM of the intensity map indicates limits of the source size that can be coupled through the FTS. 

\begin{figure}[bht]
  \centering
  \includegraphics[width=1\linewidth]{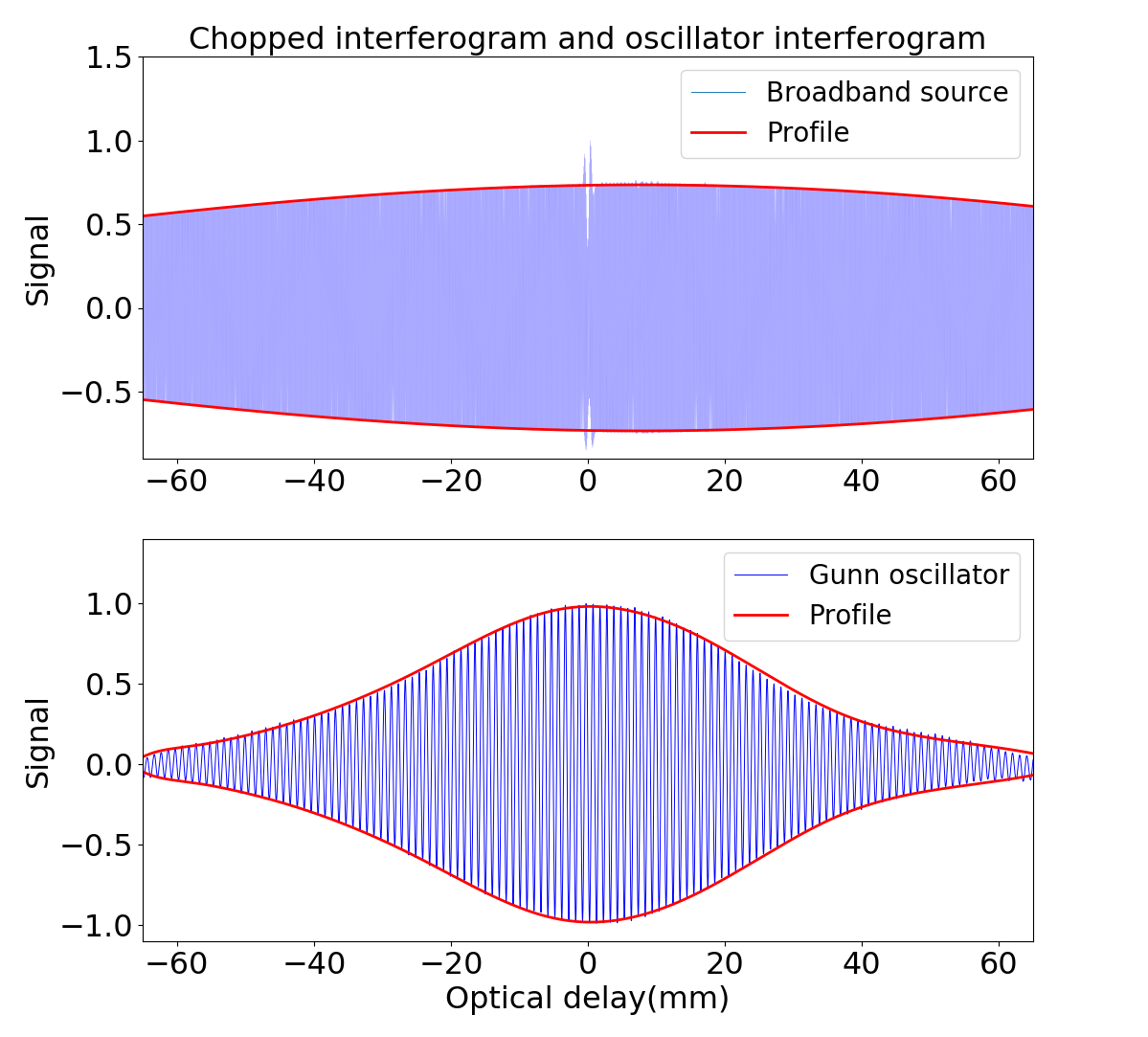}
  \caption{\textit{Top}: Delay-dependent transfer efficiency. The blue region is unresolved oscillating chopper modulation, the envelope of which is proportional to the transfer efficiency.   The envelope is obtained by fitting a spline to the maxima(and the minima) of the oscillating pattern. Note that interference is only within $\pm$3~mm of optical delay for the broadband source and is excluded when fitting for the envelope. The transfer efficiency decreases as the mirror moves away from zero optical delay because less beam is captured.  \textit{Bottom}: Interferogram for an unmodulated 295~GHz Gunn oscillator. It uses the same data as  Fig. \ref{interferograms} \textit{bottom}, but over a larger range of optical delay. The oscillation is the interference pattern and is not from chopper modulation. The delay-dependent contrast loss and the transfer efficiency loss together make the fit spline decay faster than in the top figure. The shape is expected since we designed for zero contrast at 330~GHz (Section \ref{section:design}\ref{section:geometry}).}
  \label{beamloss}
\end{figure}

\begin{figure}[bht]
  \centering
  \includegraphics[width=1\linewidth]{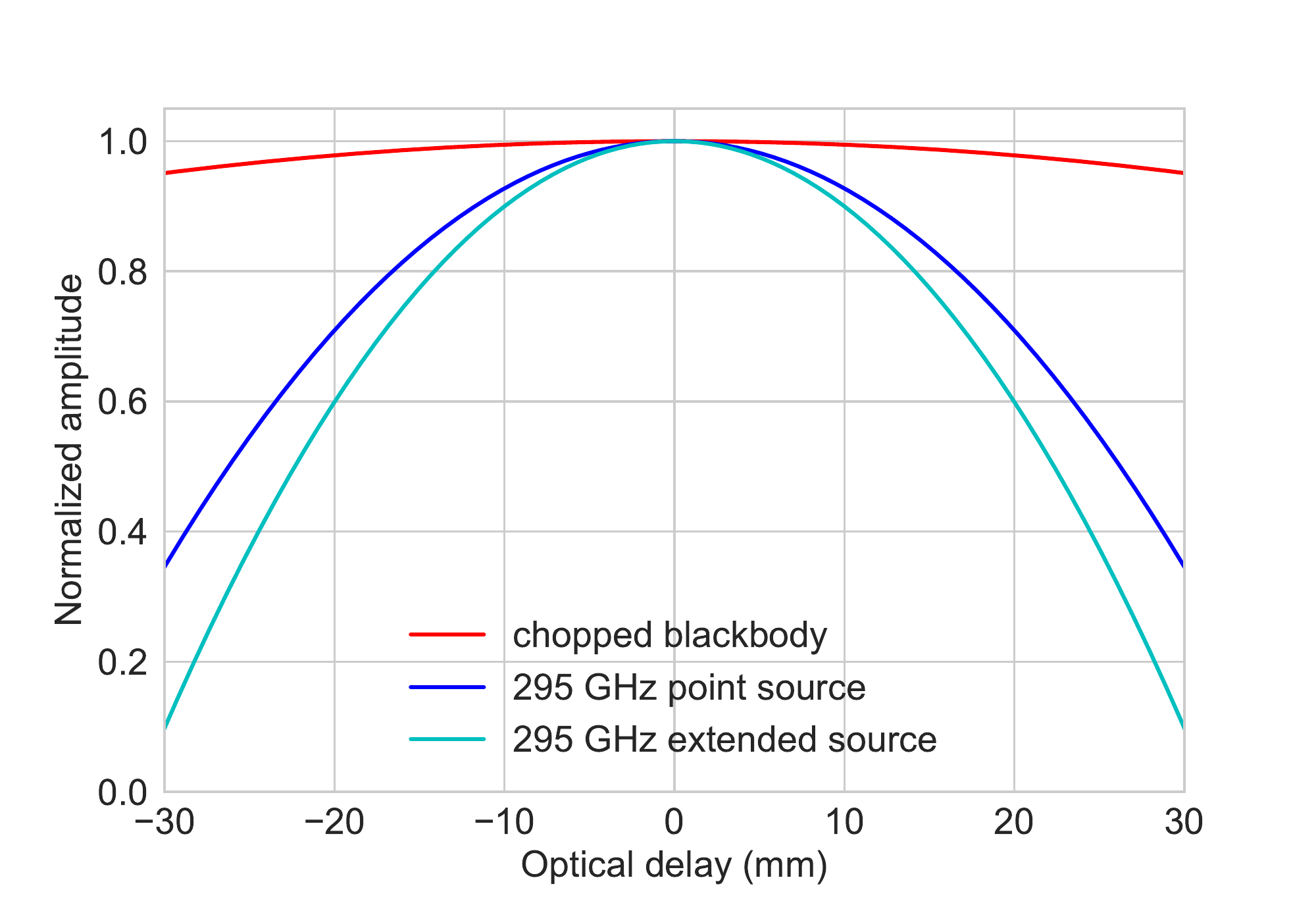}
\caption{A comparison of the delay-dependent transfer efficiency (from a chopped blackbody), the delay window functions for a 295~GHz point source (Gunn oscillator), and the delay window functions for a 295~GHz extended source.  The window functions were found using only the central portion of the data.  We took the maxima of the oscillations in the interferogram and fitted them to a second-order polynomial.  These parabolas were then normalized to one. The envelope of the extended source falls off the most quickly due to the combined effects of transfer efficiency loss, contrast loss, and frequency shift.  The point source experiences both transfer efficiency and contrast loss. The blackbody falls off most slowly due to transfer efficiency loss alone.}
  \label{distortion_decoherence}
\end{figure} 

\subsubsection{Frequency resolution and instrument window function}
\label{section:decoherence}

The resolution of the FTS is the full width at half maximum (FWHM) of the delay window function's FFT. The delay window function is defined as the envelope of a spectrally unresolved source's interferogram (see Fig. \ref{beamloss} \textit{bottom}). Several factors contribute to limiting the width of the delay window function. The first is clearly the range of the interferogram, i.e., the maximum optical delay when the mirror is at its maximum displacement \cite{Persky95}.  The design resolution of 1~GHz would require a maximum optical delay of 182~mm and a flat delay window function. However, the actual delay window function is not flat and is effectively narrower than this due to the effects outlined in the previous sections: the delay-dependent transfer efficiency, the reduce of contrast (decoherence) at large optical delays, and the averaging of the frequency shift due to the effects of an extended source. These non-idealities made the achieved resolution worse than the designed 1~GHz. The maximum optical delay of the hardware just needs to be larger than the width of the delay window function with these effects and was made to be 87~mm.

The top panel of Fig.~\ref{beamloss} is the response vs. delay for a slow scan with a chopped broadband source (top panel of Fig.~\ref{spectra}) over the full range of mirror motion. We fit the local maxima of the oscillating curve to a spline to get the envelope, which is proportional to the delay-dependent transfer efficiency. The bottom panel is an unmodulated interferogram of a 295~GHz source with a narrower envelope due to delay-dependent decoherence. Fig.~\ref{distortion_decoherence} shows the fit envelope of these two situations as well as a third which is the result of combining the frequency shifts for varying parts of the source focal plane outlined in Section \ref{section:measurements}\ref{section:beam_pattern}. The FFT FWHMs of the red, blue, and cyan envelopes in Fig.~\ref{distortion_decoherence} are 2, 4, and 6~GHz respectively, which are the frequency resolutions for the ideal case with no decoherence, a 295~GHz point source with decoherence, and an extended 295~GHz source with decoherence.

\section{Ray-trace Simulations}
\label{section:simulation}
\subsection{Ray Trace}
\label{section:raytrace} 

\begin{figure}[!ht]
  \centering
  \includegraphics[width=1\linewidth]{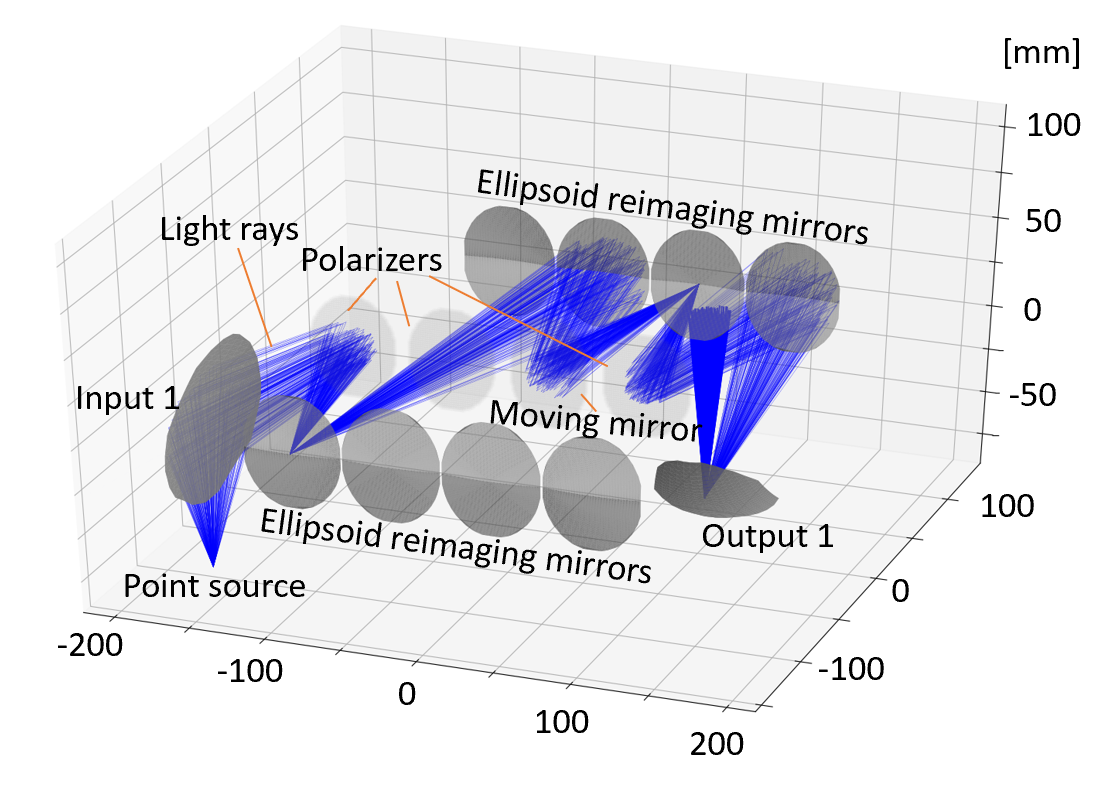}
  \caption{An example ray-trace with rays starting at a point on the source plane in the lower left. The polarizers and moving central mirror are shown in light grey along the center line of the FTS while the mirrors on the outside of the box are dark grey. This image is rotated 180 degrees from Fig.~\ref{fts_picture} about the $z$ (vertical) axis. All rays in this image are linearly polarized so they all reflect from polarizer A. Rays reflecting from polarizer B are not shown for clarity. To illustrate the optics, all rays are emitted from a single point in the source plane. Because each mirror images the previous optical element onto the next one, every other mirror is struck at a single point, as described in Section \ref{section:design}\ref{section:overview}. As can be seen, in this geometry, the detector plane above the instrument, is illuminated with an image of the first mirror.}
  \label{raytracing}
\end{figure}

To better understand our measurements in Sec~\ref{section:measurements}, we created a modified ray-trace simulation of the FTS, which is detailed in a companion paper \cite{liu19}. It also offers an improved way to balance the instrument non-idealites with constraints of size and instrument requirements. In the simulation, each ray is sourced at the focal plane of the first input mirror. The spatial and angular density of the light rays follows the beam pattern of the emitting source. Each ray has the complex phase and polarization information tracked as it passes between the mirrors and polarizers. All paths for each ray is followed to the plane of the focus of the last output optic (or lost). At that plane where the detector is modeled as a perfect power absorber, the power is calculated taking into account the relative phase of the arriving rays. Fig. \ref{raytracing} shows an example where the rays from a point source are transferred through the FTS along one of the interfering light paths. 

After verifying the simulation by comparison with measurement, it provides flexibility in exploring alignment errors and further optimizations because it is easy to change the sizes, positions, and alignment angles of the optical elements to test the construction requirements for the design limits.  The positions of the light rays enable us to see how much light rays spill over the edges of optical elements, which is related to transfer efficiency and beam loss discussed above. The distances that the light rays traveled are proportional to the delay and can be used to construct an accurate relationship between the optical delay and mirror position. Using this, we can correct for the frequency shift caused by assuming $d=4y\cos{\phi} $ for all the light rays within the \'etendu. The phase of the light rays can help model the interference, and the polarization information can help study effects caused by imperfect polarizer angles. The ray trace has limitations: it cannot model diffraction and decoherence. The ray trace assumes we know the accurate position of each light ray, but due to diffraction, every photon's wave function spread over the surface area of all optical elements and thus has a finite-sized diffraction pattern on the output focal plane. As the center mirror moves, the diffraction patterns for different interfering paths have less overlap and become more different in their phases, resulting in a reduced coherence level. The right way to model diffraction and decoherence is to do diffraction propagations by doing Fourier Transforms for the aperture fields, which is beyond the scope of our simulation. In the subsections below we highlight some applications of our simulation that can be done.

\subsection{frequency shift simulation}
\label{section:distortion}

The frequency shift results from the discrepancy between the optical delay we use in our analysis and the actual optical delay generated by the FTS box, which can be simulated by tracking the distance each light ray has traveled. In the simulation, we use a point source with the appropriate beam shape to approximate the Gunn oscillator in Section \ref{section:measurements}\ref{section:beam_pattern}. We move the source across the input focal plane to see how optical delay changes with source location. The ratio between the optical delay we assumed in the analysis and the actual optical delay from the simulation is the inverse of the ratio between the analyzed frequency and the actual frequency. 
The simulation shows that the alignment of the polarizers and mirrors needs to be better than 0.3~deg for the optical delay shift or frequency shift to be within 1\%. Our prototype FTS did not have tolerances at this level, however this can be easily improved.  The simulation doesn't have the exact same configuration as the hardware and generates a frequency shift pattern different from Fig. \ref{beammap}. An alignment accuracy of 0.3~deg is unnecessary for our current applications (detector and material characterizations). If such accuracy is needed, the wire grid polarizer mounts will need to be made more precisely. The polarizers now sit on the the bottom wire-gluing layers, which have non-uniform thicknesses. With better mounts and alignments, we will be able to use the simulation to correct for the frequency shift.

\subsection{Simulated transfer efficiency loss}
\begin{figure}[!ht]
  \centering
  \includegraphics[width=1\linewidth]{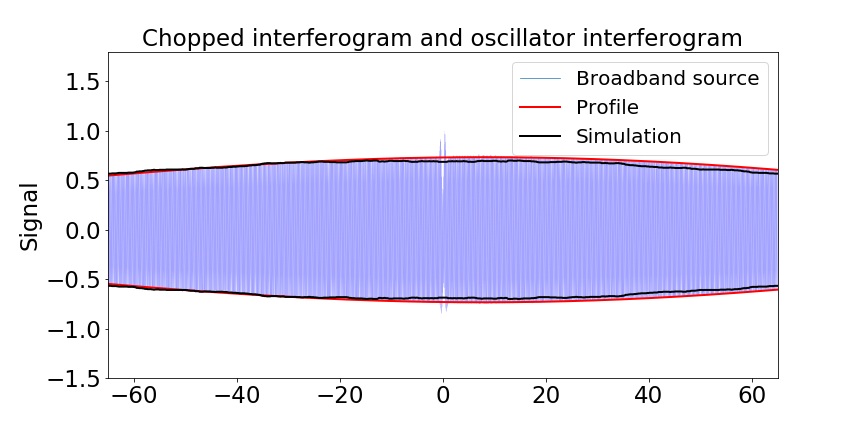}
  \caption{The simulated transfer efficiency loss (the black lines) is compared to the experimental result (the red lines). The measured transfer efficiency loss is the envelope of the modulated interferogram (the same as in Fig.\ref{beamloss} \textit{top}). }
  \label{simbeamloss}
\end{figure}

To simulate the transfer efficiency loss shown in Fig.\ref{beamloss} \textit{top}, we track the position of all the light rays and remove the rays spilling over the edges of the optical elements. The percentage of simulated rays that can reach the detector as a function of optical delay generates a shape of transfer efficiency loss. The simulated result is compared to the measurement in Fig.\ref{simbeamloss}, and they match well. The small difference between the simulation and measurement could be caused by the alignment errors of the optical elements compared to the ideal hardware configuration in the simulation.

\section{Applications}
\label{section:applications}
\begin{figure}[!ht]
  \centering
  \includegraphics[width=.95\linewidth]{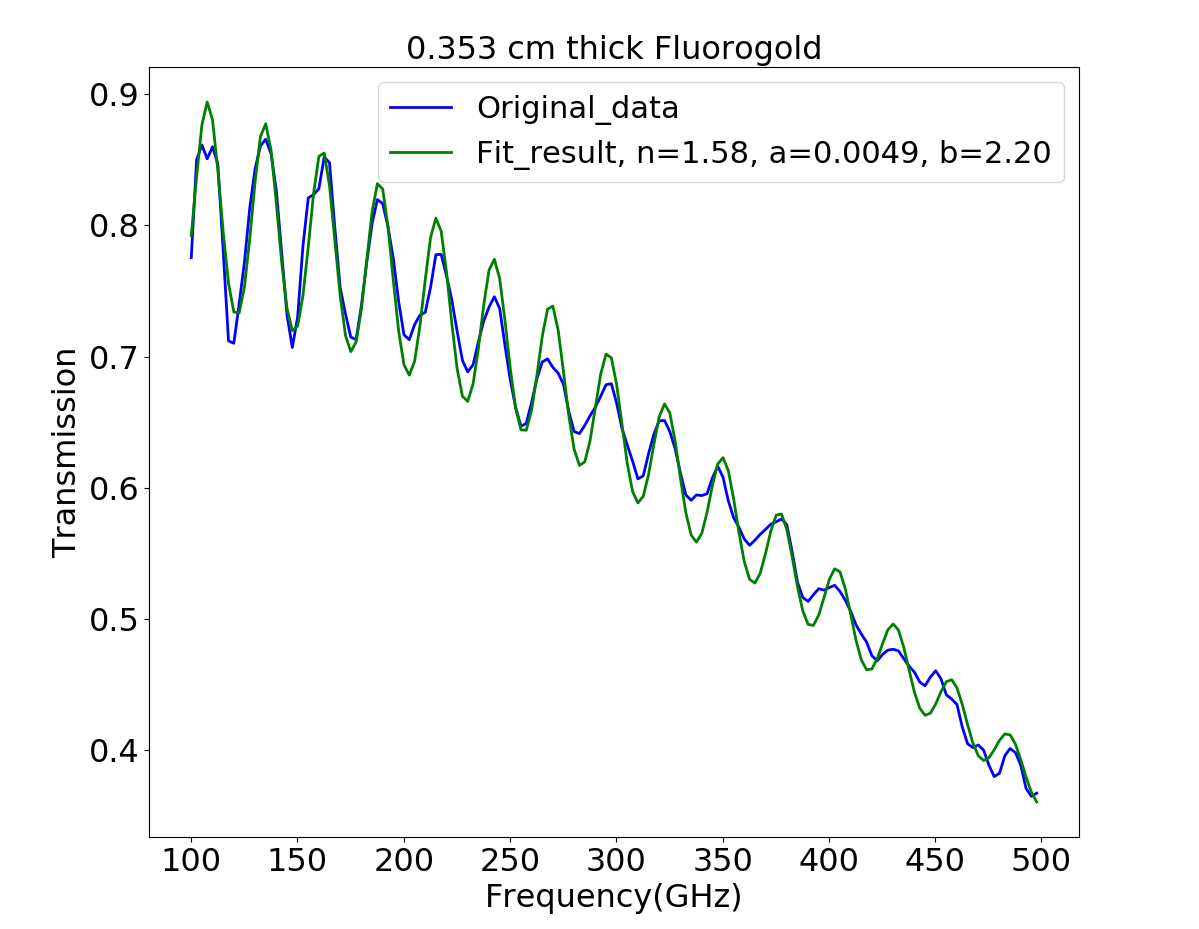}
  \caption{Measured spectrum for a piece of 0.139~in-thick fluorogold. The refraction index can be extracted from the fringing pattern, and the frequency-dependent loss parameters $a$ and $b$ can be fit from the decaying speed. The fit curve is compared to the data. }
  \label{fluorogold}
\end{figure}

The FTS in this paper has a broad range of applications in millimeter-wave astronomy and cosmology. It can be used to measure the transmission properties of optical materials at millimeter wavelengths. To measure the transmission spectrum, two measurements need to be taken: one with the optical material in the optical path and one without. Dividing these two measurements gives the absolute transmission of the optical material. Fig.\ref{fluorogold}  is a sample measurement for Fluorogold (a glass-filled Teflon) \cite{fluorogold}, which is fit to a model with the loss parameters a and b (defined in \cite{halpern86}) and the refraction index of the material. The fit parameters agree with \cite{halpern86} within 10\%. The difference can be attributed to the difference of fiber alignment directions in our Fluorogold sample and the sample used in  \cite{halpern86}. The fringe contrast at higher frequencies is less than at lower frequencies due to the degradation of frequency resolution at higher frequencies.

\begin{figure}[!ht]
  \centering
  \includegraphics[width=.95\linewidth]{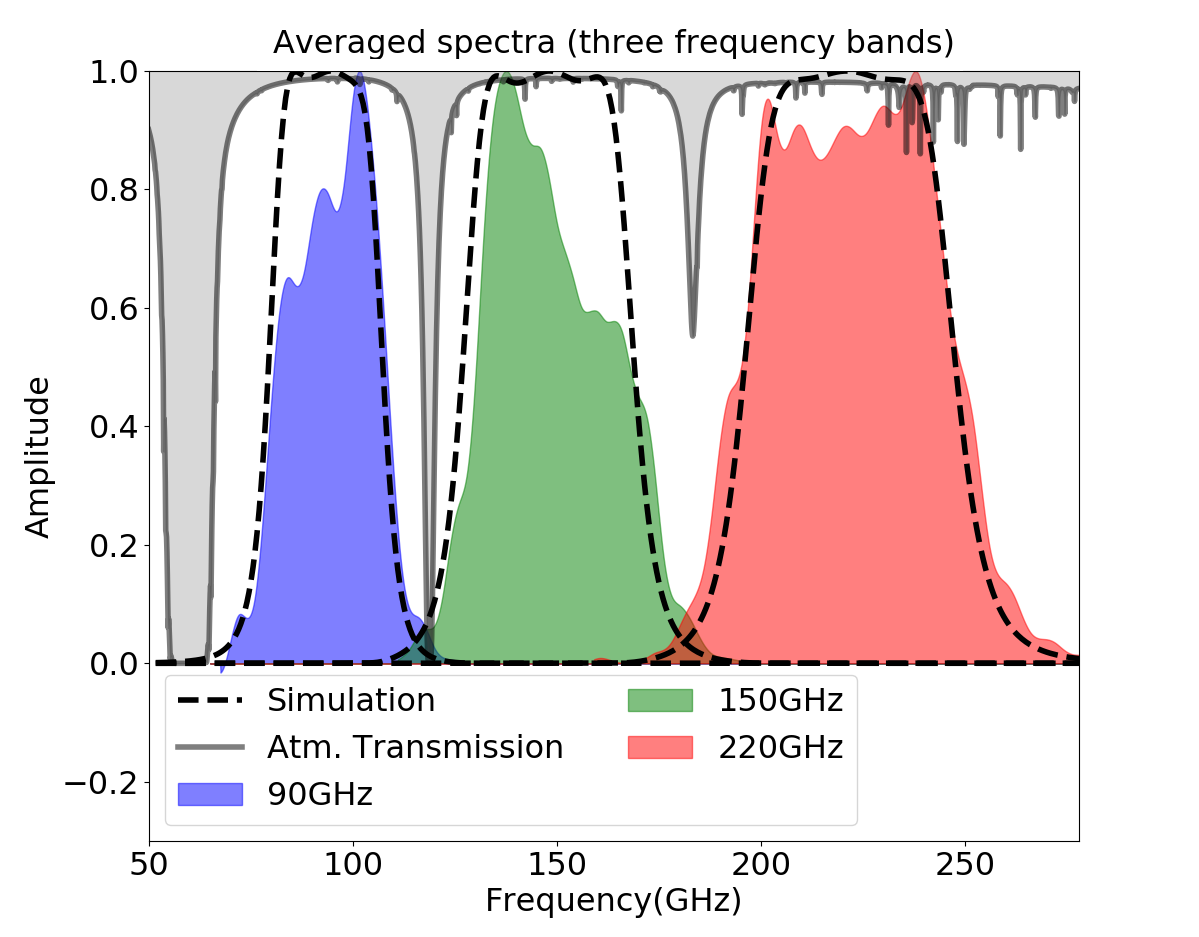}
  \caption{The measured spectral bands for the three different types of detectors used by SPT-3G. The bands are normalized to one. The atmospheric transmission and the simulated bands are also plotted. The bands lie within the atmospheric transmission window and agree with the simulations in band edges within 1\%. Band tops are different from simulations due to other optical elements not included in the simulation. }
  \label{sptbands}
\end{figure}

It can also be used to probe a detector's sensitive frequency band, which was demonstrated by the SPT-3G collaboration \cite{pan18}. In this test, we used a blackbody with a known spectrum as the source and the detector to be tested for measuring the spectrum. The measured spectrum is the product of the detector's spectral response, transmission of optical elements in the optical path, and the source's frequency spectrum, which is known and can be divided out.   Fig. \ref{sptbands}  shows a sample measurement of the SPT-3G  detectors' frequency bands. The detectors are designed to have frequency bands centered on 95, 150, and 220~GHz. The measured frequency band edges agree with the simulation within 1\%. The tops of the bands have some discrepancies from the simulation due to other optical elements in the optical path not included in the simulation, like lenses and their anti-reflecting coatings. The FTS was automated to measure SPT-3G's detector bands at a speed of 250 detectors/hour, so it's ideal for systems with a large number of detectors, especially future CMB experiments like CMB-S4 \cite{abitol17}. CMB-S4 requires the frequency accuracy of the bandpass to be within 1\% \cite{ward18}, which can be achieved by the FTS with frequency calibration by a Gunn oscillator. The FTS was also used to cross-check the spectral response of detectors for the KECK experiment \cite{staniszewski2012keck}.  The measurements were found to agree well with independent results measured using an alternative Michelson FTS design (T. Germaine, private comm.).                     

Another application is to measure sources with unknown spectra, like the CMB spectrum. For CMB spectrum measurement,  a well-calibrated blackbody can be placed at Input 2 to differentiate the input CMB signal at Input 1. This design allows us to measure small deviations of the CMB spectrum from a perfect blackbody (spectral distortion of the CMB). Both outputs can be used for systematics control.

\section{Conclusions}
\label{section:conclusion}
In this paper, we discussed the design and development of an FTS for millimeter wavelengths. The design has a small size of $355\times 260\times 64$~mm, which is the minimum size that can achieve the designed throughput, resolution, and frequency range. The size parameters are functions of the design requirements, so better design specifications would require a larger instrument. We did comprehensive measurements for the FTS in different configurations to verify that it meets the design needs. The transfer efficiency and the contrast are 92\%$\pm$5\% and $-$55\%$\pm$3\%, respectively. The frequency resolution of the FTS for a point source is 4~GHz at 295~GHz and is worse than the design target of 1~GHz due to reduced interference intensity at larger optical delays, especially at higher source frequencies. For an extended source, the weighted frequency shift is within $\pm$0.1\%. For a point source, the measured frequency can shift by $\pm 2.5\%$ as a function of the source's location on the focal plane.   To constrain the specifications of the instrument, understand the measurements, and potentially correct for some of the non-idealities in the measurements, we have developed a ray trace simulation software for our system. The simulation method can be used for other FTS systems as well. Our FTS has a broad range of applications, including characterizing material properties, detector performance, and spectra for unknown sources like the CMB. 

\section*{Funding}
\label{section:Funding}
National Science Foundation (PLR-1248097); NSF Physics Frontier Center grant (PHY-1125897); the Kavli Foundation; the Gordon and Betty Moore Foundation (GBMF 947). 

\section*{Acknowledgements}
\label{section:acknowledgement}
The authors thank John Carlstrom for his instrument and useful discussions.

\newpage
\bibliography{references}

\bibliographyfullrefs{references}

\end{document}